\documentstyle[12pt,aaspp4]{article}
\def\Ginga{\hbox{\it Ginga}}
\tighten

\lefthead{Band et al.}
\righthead{GRB Line Candidates}

\begin{document}

\title{BATSE GAMMA-RAY BURST LINE SEARCH: \\
IV. LINE CANDIDATES FROM THE VISUAL SEARCH}
\author{D. L. Band, S. Ryder, L. A. Ford, J. L. Matteson}
\affil{CASS 0111, University of California, San Diego, La Jolla, CA  92093}
\author{D. M. Palmer, B. J. Teegarden}
\affil{NASA/GSFC, Code 661, Greenbelt, MD 20770}
\author{M. S. Briggs, W. S. Paciesas, G. N. Pendleton, R. D. Preece}
\affil{University of Alabama at Huntsville, Huntsville, AL 35899}
\centerline{\it Received 1995 July 11; accepted 1995 August 22}
\centerline{To appear in the February 20, 1996, issue of}
\centerline{{\it The Astrophysical Journal}}
\begin{abstract}
We evaluate the significance of the line candidates identified by a visual
search of burst spectra from BATSE's Spectroscopy Detectors. None of the
candidates satisfy our detection criteria:  an $F$-test probability less than
$10^{-4}$ for a feature in one detector and consistency among the detectors
which viewed the burst.  Most of the candidates are not very significant, and
are likely to be fluctuations. Because of the expectation of finding absorption
lines, the search was biased towards absorption features. We do not have a
quantitative measure of the completeness of the search which would enable a
comparison with previous missions. Therefore a more objective computerized
search has begun.
\end{abstract}
\section{INTRODUCTION}
The visual search for absorption lines in the gamma ray burst spectra
accumulated by the Burst and Transient Source Experiment (BATSE) on the {\it
Compton Gamma Ray Observatory (CGRO)} has resulted in a number of line
candidates, but no definitive detections (Palmer et al. 1994, hereafter
Paper~I).  Here we present the results of a systematic evaluation of the line
candidates identified thus far, and discuss the two most significant
candidates. The existence of lines in the BATSE spectra is one of the major
unresolved observational issues confronting BATSE since the absorption lines
reported by the KONUS (Mazets et al. 1981), {\it HEAO~1} (Hueter 1987) and {\it
Ginga} (Murakami et al. 1988) detectors were attributed to cyclotron absorption
in $\sim$10$^{12}$ gauss magnetic fields (\cite{wang89}), which suggests that
bursts originate on neutron stars, the only known anchors for teragauss fields.

This paper is part of a series describing and analyzing our search for lines in
the BATSE data.  Paper~I reported the absence of line detections by the visual
search of the BATSE spectra.  Band et al. (1994, hereafter Paper~II) developed
the methodology for comparing the absence of a BATSE detection with the
detections reported by previous missions; a preliminary calculation shows that
the apparent discrepancy between BATSE and {\it Ginga} is not yet compelling.
Band et al. (1995, hereafter Paper~III) demonstrated that BATSE would have
detected the lines that {\it Ginga} observed; the simulations in Paper~III also
provide the probabilities for detecting lines in the BATSE spectra which are
necessary for a comprehensive comparison between BATSE and previous detectors.
The visual search provides neither the statistics necessary for this
comprehensive search, nor measures of its completeness, as is evident from the
candidate evaluation presented here.  Therefore, we have begun a more
sophisticated computerized search (Schaefer et al. 1994).

The evaluation of line candidates involves many instrumental and data analysis
details which are summarized in \S 2.  We first present the results of the
candidate evaluation (\S 3), and then discuss two particular candidates in
greater detail (\S 3.1-2).

\section{METHODOLOGY}

We search for line features in the spectra accumulated by BATSE's Spectroscopy
Detectors (SDs); the SDs have been described in detail elsewhere  (Fishman et
al. 1989; Band et al. 1992).  In brief, the SDs are simple 5" diameter by 3"
thick cylindrical NaI(Tl) scintillation detectors with an energy resolution of
19\% at 60~keV. An SD is found in each of BATSE's eight modules located at the
corners of the {\it CGRO} spacecraft.  Each module also contains a Large Area
Detector (LAD) for detecting transient events, determining their positions, and
recording their time histories.  For 4-10 minutes after a burst trigger, BATSE
accumulates 192 SD spectra with 256 channels spanning two energy decades with
a
low energy cutoff between 15 and 30~keV (the SHERB data type). These spectra
are gathered from four detectors, with a greater number from the most brightly
illuminated detectors. The accumulation times are based on the count rate.
Background spectra are accumulated every $\sim5$ minutes when BATSE is not in
burst mode (the SHER data type).

Unfortunately, a few channels just above each spectrum's low energy cutoff are
distorted by the SLED ({\bf S}D {\bf L}ow {\bf E}nergy {\bf D}istortion), an
electronic artifact discovered after launch (Band et al. 1992).  For a typical
SD in its highest gain setting the upper end of the SLED is between 15 and
20~keV. Although the SLED can be partially mitigated by the calibration
software, the channels affected by the SLED will probably never be trusted for
more than determining the continuum below a line candidate.  However, for the
purpose of establishing consistency among all detectors we do search for
features in the SLED-affected channels when lines have been found at the same
energy in other detectors.

For an observed spectral feature to be considered intrinsic to the incident
spectrum (i.e., a ``real'' feature of astrophysical interest), the probability
that it is a statistical fluctuation must be small, and the spectra accumulated
by all detectors must be consistent.  Thus the first detection criterion is
that the feature is significant in at least one detector, while the second
criterion is consistency among all detectors which could have observed the
feature.  Note that we do not require multiple independent detections (although
it is very rare that a feature would be observable by only one detector), nor
must a feature be significant in all detectors where it is observed.  As we
showed in Paper~III, fluctuations can vary the apparent significance of an
absorption line.  In addition, the probability that a line will be significant
enough to be considered a detection varies with the burst angle (the angle
between the burst and the detector normal).  Thus we cannot expect significant
detections in all detectors.

We use the $F$-test to calculate the probability that an observed feature is a
fluctuation.  As applied here, the $F$-test compares a spectral fit with a
continuum-only model to the fit with a continuum+line model.  Assume the
continuum-only fit gives $\chi_1^2$ with $\nu_1$ degrees-of-freedom, while the
continuum+line fit results in $\chi_2^2$ with $\nu_2$.  We define the $F$
statistic as (Martin 1971)
\begin{equation}
F = \left({{\chi_1^2-\chi_2^2}\over{\nu_1-\nu_2}}\right)
   \left({{\nu_2}\over{\chi_2^2}}\right) \quad .
\end{equation}
This $F$ statistic is characterized by a normalized distribution which is a
function of $\nu_1-\nu_2$ and $\nu_2$; the integral over this distribution from
$F$ to infinity is the $F$-test probability $P(\ge F)$, which we define as the
significance of the feature, with a small probability indicating a more
significant feature.  We choose a maximum probability $P(\ge F)$ as the
threshold for a candidate to be considered a detection.  Note that $P(\ge F)$
is only the {\it a priori} probability of a fluctuation with the observed
parameters;
it does not consider all the possible ``trials'', the range of parameters
(e.g., time intervals, line centroids, widths, etc.) in which line-like
fluctuations could have occurred.  If there are a large number of trials then
it is likely that in some trial there will be a rare fluctuation. The
probability that a fluctuation will produce a spurious line feature with a
given significance somewhere in our dataset is the product of $P(\ge F)$ and
the number of trials; therefore we require a very small $P(\ge F)$ to conclude
the line is real.  The calculation of the number of trials in our data is a
difficult issue in standard ``frequentist'' statistics which we will address
later in this series of papers.  We estimate there are of order $\sim 3000$
trials in our spectra, and therefore use a detection threshold of $P(\ge F)\le
10^{-4}$.  While this threshold may not appear stringent enough since the
probability of a false positive is then $\sim30\%$, we also require consistency
among detectors (Paper~I).  The expected distribution of fluctuations will be
analyzed in greater detail later in this series of papers.

Only $\sim 20\%$ of the bursts detected by BATSE have a sufficiently high count
rate to warrant detailed spectral analysis, which roughly corresponds to a LAD
peak count rate over 10,000 counts-s$^{-1}$ in the 50-300~keV band (i.e., LAD
discriminators 2 and 3).  Thus of the over 1000 bursts detected, only $\sim$250
have been searched, and of these bursts only $\sim$50 were intense enough for
lines comparable to those observed by \Ginga\ to be detectable.

Although the spacecraft telemetry provides SHERB spectra for the four most
brightly illuminated detectors, not all the data are appropriate for line
searches.  First, the detectors must be run at high gain to extend the
accumulated spectra low enough to cover the energy range where previous
missions reported spectral features. For most of the mission, two of the eight
SDs have been operated at low gain (to satisfy other mission objectives), and
the gain of one of the nominally high gain detectors cannot be pushed high
enough to cover energies below $\sim30$~keV.  Second, the burst angle may be
greater than $\sim 80^\circ$, the angle at which the spacecraft and the rest of
the BATSE module affects the incident flux and our detector response model
becomes unreliable. The SHERB data are from the SDs in the modules of the four
most brightly-illuminated LADs; these LADs normally have burst angles less than
$90^\circ$.  However, the detector normals of the SDs and LADs are offset by
$18.5^\circ$, and therefore a module whose LAD has a burst angle less than
$90^\circ$ may have an SD with an angle greater than $90^\circ$. Earth-scatter
of GRB photons off the atmosphere only complicates matters by occasionally
increasing the count rate in detectors with large burst angles. Thus usable
spectra may be available from only two or three SDs.

The visual search inspects background-subtracted spectra on different time
scales.  By adding together the basic SHERB spectra, we construct spectra for
the entire burst and for the major temporal substructures (e.g., each major
emission spike). In addition, spectra are produced on the shortest time scale
over which all the detectors are accumulated, that of the third and fourth rank
detectors (i.e., the SDs in the third and fourth most brightly illuminated
modules). Thus spectra are available from all relevant detectors over the same
time ranges, allowing comparisons on the same time scale. The resulting
background-subtracted spectra are inspected visually for both emission and
absorption lines over the entire energy range.  Features which appear to be
very significant are studied further by inspecting spectra averaged over
different time ranges.  Thus a feature might be identified in a spectrum
averaged over the third rank accumulation time (e.g., from 3.328 to 6.592~s
after the trigger) but might be most significant in a spectrum averaged over an
overlapping time range (e.g., from 4.160 to 7.360~s). This search is sensitive
to features which are significant on the searched time scales, or which are
apparent to the eye in the searched time intervals but become more significant
by varying the time range; thus this search is not sensitive to short lived
features which are washed out in spectra accumulated over longer periods, nor
to features which persist a long time but are not evident in the searched
spectra over shorter time periods. Our simulations indicate that the eye is
sensitive to features with an $F$-test significance of $10^{-2}$ or smaller
(i.e., $P(\ge F)\le 10^{-2}$).  Since in this study we find that we identified
many candidates with significances of order 0.1, we should have detected
features which persisted over many of the searched spectra.

We evaluate the line candidates identified by the visual search. Because the
gain differs between detectors and varies with time, each burst's spectra must
be calibrated for each detector (Band et al. 1992). Our response matrices
include Earth-scatter.  Background spectra are created channel-by-channel by
fitting 2nd, 3rd, or 4th order polynomials to spectra before and after the
burst.  As always, background creation is something of an art; we evaluate the
calculated background by the quality of the fits to each channel. These fits
provide the background as a function of time; for background subtraction these
functions are integrated over the time interval of interest.

We fit the spectra with the standard Marquardt-Levenberg algorithm minimizing
$\chi^2$ (Bevington 1969, pp.~232-241; Press et al. 1992, pp.~678-683).  In
brief, the parameters of a model photon spectrum are varied to minimize the
difference---quantified by $\chi^2$---between the calculated count spectrum
(the model photon spectrum folded through the detector response) and the
observed spectrum.  We define $\chi^2$ with model variances, requiring
additional terms in the gradients used by the Marquardt-Levenberg algorithm
(see the appendix of Ford et al. 1995).

The continuum is usually fit with the four parameter ``GRB'' (for lack of a
better name) model (Band et al. 1993),
\begin{eqnarray}
N_C(E) =& A (E/100)^\alpha
   \exp\left[-\left({E\over{E_0}}\right)\right]\quad , \quad
   &E\le (\alpha-\beta)E_0 \\
=& A^\prime (E/100)^\beta \qquad , \quad &E> (\alpha-\beta)E_0 \quad
,\nonumber
\end{eqnarray}
where $A^\prime$ is chosen so that $N_C(E)$ is continuous at
$E=(\alpha-\beta)E_0$.  In a few cases where we find $E_0$ to be very large or
$\beta$ is nearly equal to $\alpha$, we simplify the continuum model, either by
using a simple power law or by dropping the high energy power law in eq.~2 (the
COMP model); the decision to simplify the continuum can be justified in terms
of the effect on the reduced $\chi^2$.  An apparent absorption line is usually
modeled as a multiplicative line,
\begin{equation}
F(E)=N_C(E)\, \exp\left(-\phi\left(E\right)\right) \quad ,
\end{equation}
and an emission line as an additive line,
\begin{eqnarray}
F(E)=N_C(E) + \phi(E) \quad ,
\end{eqnarray}
where $\phi(E)$ is the Gaussian line profile (with three parameters: amplitude,
line centroid and line width).  For absorption lines whose intrinsic width
appears to be much narrower than the instrumental resolution we also use a
simple two parameter ``black'' line model with no flux over an energy range
centered on the line centroid; for very narrow lines the instrument is unable
to resolve the line profile.

We always fit the spectra over the greatest energy range possible.  The low
energy cutoff is chosen to avoid the SLED while the highest available energy
channel is included.  In Paper~III we found that extending the fitted energy
range to energies much greater than the features of interest increased the
significance of the features, in part because the continuum is better
determined.  On the other hand, we were concerned that the choice of the low
energy cutoff $E_1$ could make a large difference in the calculated
significances.  The exact value of $E_1$ which should be used is not automatic
because the upper end of the SLED artifact is not clear-cut, and analysts may
choose different values (in particular there is a tendency to choose round
numbers, e.g., 20~keV when the SLED ends at 17~keV).  To test the effect of
varying $E_1$, we calculated the significance of the 54~keV candidate
absorption line in GB~920307 using different values of $E_1$, as is shown by
Figure~1.  As can be seen, there is not a meaningful change in significance for
$E_1$ between 15 and 30~keV, and the significance decreases ($P(\ge F)$
increases) only as $E_1$ approaches the candidate line energy.  Thus small
changes in $E_1$ should not make a large difference.  Of course, the
significance will be sensitive to the choice of $E_1$ if the line energy is
near the low energy cutoff.
\section{RESULTS}
Table~1 describes the candidates and the fits to them, while Figure~2 presents
the distribution of their significances.  Few candidates are more significant
than $P(\ge F)\le 0.01$, and none exceeds our detection threshold of $P(\ge F)
= 10^{-4}$, although two nearly satisfy the significance criterion; these two
are discussed in greater detail below.

The significance distribution of the line candidates in Figure~2 is shaped by a
number of components.  First are the inevitable line-like statistical
fluctuations, with many low significance fluctuations and few at the high end
(i.e., small values of $P(\ge F)$).  The normalization of this distribution is
proportional to the number of trials in the burst dataset, a poorly determined
quantity.  Second is the distribution produced by real spectral lines (if lines
exist).  These lines are characterized by a (currently unknown) distribution of
parameters.  Paper~III showed that as a result of the fluctuations in any given
observation, the same line may be observed by identical detectors with very
different significances. Thus the observed significance distribution is the
convolution of the line parameter distribution and the distribution of
significances for a given set of line parameters. Note that it is entirely
possible that many of our low significance (large $P(\ge F)$ value) candidates
which do not satisfy our detection criteria may be real lines. Finally,
incompleteness truncates the low significance end of the observed significance
distribution; most candidates with $P(\ge F)\sim 0.1$ will not be identified by
the search.

We suspect that most or all of our line candidates are fluctuations. Our
intention was to identify those features which appeared to have $P(\ge F)\le
0.01$; the large number of candidates with $P(\ge F)\ge 0.1$ leads us to
suspect the accuracy of the qualitative (i.e., ``eyeball'') assessment of the
significance of a line.  We would expect equal numbers of emission line-like
and absorption line-like fluctuations.  However, absorption features were more
readily recorded because of the previous reports of absorption lines in the
15-100~keV range (as discussed in Paper~II), although emission features up to
511~keV were also flagged (emission lines have also been reported by previous
missions).  In addition, absorption lines may be more apparent to the eye than
emission lines of comparable strength because of the curvature of the burst
count spectra. Thus we find that the number of absorption line candidates (67)
exceeds that of emission line candidates (13).

The visual search does not provide the statistics of the lines which could have
been detected in the searched spectra.  These statistics are necessary to
compare different missions (e.g., BATSE with \Ginga) and to place constraints
on the frequency with which different types of lines occur (Paper~II).  In
addition, the search does not produce measures of its completeness (e.g., the
fraction of features with $P(\ge F)\sim 10^{-2}$ which were identified by the
search).  Since the visual search relies on human judgement which can be
affected by fatigue and distractions, there is always the fear that an
interesting candidate may have been missed. While we are confident that we
would have detected features with $P(\ge F)\le 10^{-4}$ in the inspected
spectra, we are relying on identifying less significant features in order to
detect lines which persist over many of the searched spectra.  Thus we cannot
conclude definitively that detectable lines do not exist in the BATSE spectra.
For these reasons we have begun a computerized search (Schaefer et al. 1994)
which is unaffected by human subjectivity and which will provide relevant
statistics.  In addition, this computerized search will consider almost all
possible spectra which can be formed from consecutive SHERB spectra.

\subsection{The Feature in GB~920315}

The most significant candidate absorption line is found in GB~920315 with
$P(\ge F) = 1.6 \times 10^{-4}$, just above our detection threshold.  The
candidate is a feature at 80~keV in a spectrum accumulated between 1.152 and
1.728~s from SD1 (i.e., the SD in module 1), which is the 2nd rank detector
with a burst angle of 58.3$^\circ$.  The 1st rank detector, SD3, was in a low
gain setting (and therefore did not cover 80~keV), and the 3rd and 4th rank
detectors, SD6 and SD7, had burst angles of 121.7$^\circ$ and 105.4$^\circ$,
respectively.  Therefore, SD1 was the only SD capable of seeing a feature below
$\sim 300$~keV, making consistency between detectors automatic; had the feature
in SD1 met our significance criterion, it would have technically been a
detection.  Features are not evident at 80~keV in LAD spectra (with poorer
spectral resolution but better statistics than SD spectra), although the
sensitivity of the LADs to lines of this type has not yet been determined. As
can be seen from the count spectrum in Figure~3, the candidate at 80~keV draws
much of its significance from one particularly low channel; however, this low
channel is surrounded by a number of channels which dip below the general
trend.

As we argued above, there is a probability of order $\sim 30$\% of a
fluctuation with a significance of $\sim 10^{-4}$ somewhere within our data
set, and therefore this feature can be attributed to a statistical fluctuation.
Indeed, we would not have been comfortable declaring this candidate a detection
had its significance been a factor of 1.6 smaller because of the absence of
confirmation by other detectors and the importance of one low point.  Our first
definitive detection must be more convincing.

\subsection{The Feature in GB~930506}

The candidate in GB~930506 has attracted a great deal of attention (Ford et al.
1994; Preece et al. 1994; Freeman \& Lamb 1994; Paper~I).  An absorption
feature was found at 55 keV in SD2; Figure~4 presents the count spectrum. The
feature has been evaluated with fits by various continuum and line models over
different energy ranges; in many cases the resulting $P(\ge F)$ satisfied our
detection criterion.

However, the feature in SD2 is absent in the spectra from other BATSE
detectors.  Two other SDs and two LADs, all with smaller burst angles than SD2,
also viewed the burst (see Table~2). Unfortunately, the SD with the smallest
burst angle (SD3) was at a low gain setting, and therefore did not cover the
energy of the line candidate. In SD7 the upper edge of the SLED fell at the
energy of the line in SD2.  Using the methodology of Band et al. (1992), Ford
et al. (1994) developed an {\it ad hoc} correction for the SLED in SD7.  With
this correction, no line is evident at 55~keV in SD7; fitting the SD7 spectrum
with a continuum+line model found a weak line at 55.9~keV with $P(\ge F)=0.73$.
The probability that the line observed in SD2 (i.e, with the parameters fit to
the SD2 spectrum) would be as nearly unobservable as the line fit to SD7 is
less than one percent; specifically, none of 200 simulated SD7 spectra using
the parameters of the SD2 fit resulted in $P(\ge F)$ as large as 0.73.
Similarly, a joint fit to SD2 and SD7 does not find a significant line feature
(Briggs 1995). No lines are evident in the two LADs with small burst angles
(Preece et al. 1994).

We reanalyzed the feature in SD2 with the procedures used to evaluate the other
line candidates; the results are presented by Table~3.  We tried three
different continuum models---a power law, COMP and GRB---and two different
line
models---multiplicative and ``black.''  The background varied significantly
over the $\sim 2000$~s before and after the burst, and there were a
number of reasonable choices of background spectra from which we could
interpolate the background at the time of the candidate. Consequently we
calculated the line significances for the six continuum+line models (three
continuum and two line models) for two different interpolated backgrounds; as
can be seen from Table~3, the resulting line significances are qualitatively
similar using the two background spectra.  From the reduced $\chi^2_\nu$ for
the different fits, also presented in Table~3, it is clear that the power law
fits are poor, while the COMP and GRB fits are acceptable ($\chi^2_\nu\sim 1$).
While the power law fits are highly significant, $P(\ge F)\sim 10^{-7}$, the
COMP and GRB fits do not meet our significance criterion for a detection.  It
is our experience that when a continuum model does not fit a spectrum well, the
line model assists in fitting the observed continuum; hence a poor continuum
model results in a significant line fit.  Thus we conclude that the high
significance of the power law fits is not an accurate representation of the
line significance.  The fitted line is narrower than the
instrumental resolution (a resolution of $\sim 11$~keV as opposed to a line
width of $\sim3.9$~keV), and therefore using the ``black'' line model, with one
fewer parameter than the multiplicative line model, is more significant by a
factor of $\sim 4$. Nonetheless, the line does not satisfy our significance
criterion for any of the acceptable line fits.

In their extensive analysis of this candidate, Freeman \& Lamb (1994) performed
fits to SD2 and joint fits to SD2+SD7, SD2+SD3, and SD2+SD3+SD7 using both the
COMP and GRB continua.  The significances of the joint fits to SD2+SD3 are
$1.6\times 10^{-5}$ and $7.4\times 10^{-5}$ for the COMP and GRB continua,
respectively.  We do not believe this affects our conclusion that the feature
is not significant.  First, the relevant significance is the largest $F$-test
probability that results from a reasonable continuum model; to show the feature
is significant we must prove that it cannot be produced by a reasonable
continuum shape.  Thus we should use the significance from the joint fit with
the GRB continuum which just barely satisfies our first criterion. Second,
adding a second detector doubles the number of degrees-of-freedom.  As
discussed in Paper~III, extending the energy range of a fit can improve the
$F$-test probability both by better determining the continuum, thereby
increasing $\Delta \chi^2$ between the continuum and continuum+line fits, and
by increasing the degrees-of-freedom. The first effect is a true reflection of
the line significance, while the second effect is an artifact of the $F$-test.
We found in Paper~III that a factor of two increase in the number of
degrees-of-freedom decreased the $F$-test probability by an order of magnitude;
therefore we conclude that most of the improvement of the significance of the
SD2+SD3 joint fit over the SD2 fit results from this degrees-of-freedom effect.

We therefore conclude that the line candidate in GB~930506 meets neither of our
detection criteria.  The $F$-test probability for the line feature in SD2 is
small but not so small as to lead us to reject the possibility that the feature
is a rare statistical fluctuation.
\section{SUMMARY}
We have evaluated the line candidates identified by the still-ongoing visual
search of burst spectra observed by the BATSE SDs.  We find that none of the
candidates satisfy our detection criteria; indeed most of the candidates are
not very significant.

Taken at face value, this evaluation of line candidates from the visual search
leaves us with no BATSE line detections, while KONUS, {\it HEAO~1} and \Ginga\
all reported detections.  However, our approximate assessment of the
consistency between BATSE and \Ginga\ (Paper~II), accurate to a factor of
$\sim$2, indicates that the apparent discrepancy between these two missions is
not yet compelling; a more definitive calculation is in progress.  Thus based
on the BATSE observations there is no reason yet to question the detections
reported by previous missions.  Clearly, we will be forced to conclude that
BATSE is discrepant with previous missions if no lines are detected after many
additional intense BATSE bursts are observed.

The visual search has not provided us with measures of its sensitivity; we do
not know what part of the space of line parameters has been probed.  In
addition, the visual search is affected by the subjectivity of the human eye,
and there is always the fear that the search may have missed a significant
line.  Consequently we have begun a new computerized search which should
overcome these limitations.  Also, this search will consider progressively
longer accumulations, thereby increasing the range of lines which can be found.
Both the ongoing visual search and this computerized search will undoubtedly
generate new candidates, and perhaps ultimately a few detections.
\acknowledgments
We thank P.~Freeman and D.~Lamb for insightful discussions regarding GB~930506.
The work of the UCSD group is supported by NASA contract NAS8-36081.

\clearpage

\addtocounter{page}{2}

\setcounter{table}{1}

\begin{table*}
\begin{center}
\smallskip
\begin{tabular}{l c c}
\tableline
\tableline
Detector & Burst Angle\tablenotemark{a} &
Energy Range (keV)\tablenotemark{b} \\
\tableline
SD3  & 43.7$^\circ$ & 250-25000 \\
SD7  & 56.2$^\circ$ & 55-4800 \\
SD2  & 73.7$^\circ$ & 15-1350 \\
LAD3 & 28.8$^\circ$ & 35-1800 \\
LAD7 & 42.8$^\circ$ & 35-1800 \\
\end{tabular}
\end{center}
\caption{Detectors viewing the GB~930506 candidate}
\tablenotetext{a}{The angle between the detector normal and the burst.}
\tablenotetext{b}{The energy range covered by the detector.}
\end{table*}
\clearpage
\begin{table*}
\begin{center}
\begin{tabular}{l c c c c c}
\tableline
\tableline
\multicolumn{2}{c}{Continuum Model} &
\multicolumn{2}{c}{Black Line\tablenotemark{a}} &
\multicolumn{2}{c}{Multiplicative Line\tablenotemark{b}} \\
Type\tablenotemark{c} & $\chi^2_\nu$  & $P(\ge F)$  & $\chi^2_\nu$  &
   $P(\ge F)$  & $\chi^2_\nu$  \\
\tableline
PL   & 1.96688 & $1.28\times10^{-7}$ & 1.72630 & $6.62\times10^{-7}$ &
1.73506 \\
     & 1.97315 & $1.04\times10^{-7}$ & 1.72861 & $2.85\times10^{-7}$ &
1.72738 \\
\tableline
COMP & 1.00083 & $2.46\times10^{-4}$ & 0.93837 & $9.30\times10^{-4}$ &
0.94320
\\
     & 0.99953 & $1.87\times10^{-4}$ & 0.93489 & $7.98\times10^{-4}$ &
0.94063 \\
\tableline
GRB  & 0.99894 & $3.20\times10^{-4}$ & 0.93850 & $1.18\times10^{-3}$ & 0.94324
\\
     & 0.99772 & $2.41\times10^{-4}$ & 0.93502 & $9.74\times10^{-4}$ &
0.94043 \\
\end{tabular}
\end{center}
\caption{Significance of GB~930506 candidate in SD2}
\tablenotetext{a}{A line model with a rectangular profile
centered on the line centroid (2 parameters).}
\tablenotetext{b}{The line model is an exponential of a Gaussian line profile
(eqn.~3---3 line parameters).}
\tablenotetext{c}{The type of continuum model:  PL---power law (2 parameters);
COMP---power law with exponential break (3 parameters); GRB---power law with
exponential break flattening to high energy power law (eqn.~2---4 parameters).
Note that PL is a subset of COMP, which in turn is a subset of GRB.}
\tablecomments{Fits to the line candidate at 55~keV in the SD2 spectrum
accumulated over 6.080-9.920~s. There are two entries for each case
corresponding to two different background spectra. The spectrum was fit over
232 channels covering 15.3--1351~keV.}
\end{table*}

\clearpage

\figcaption{Line significance as a function of low energy cutoff $E_1$
for the line candidate at 54~keV in GB~920307.}

\figcaption{Histogram of line candidate significances.  The dashed curve is the
histogram for the emission lines, while the solid curve is for all candidates.}

\figcaption{Candidate in GB~920315.  The candidate is in a spectrum accumulated
between 1.152 and 1.728~s by SD1.  The dot-dashed line indicates the best fit
GRB continuum while the solid curve shows the effect of adding a 3-parameter
line to the model. The iodine K-edge produced the dip around 35~keV. The line
appears to be formed predominantly by one low point, although there is a slight
dip in the surrounding channels.}

\figcaption{Candidate in GB~930506.  The count spectra were accumulated from
6.08 to 9.92~s after the trigger. The upper set of data points is from SD2
while the lower set (shifted down by a factor of 1/3) is from SD7, with an {\it
ad hoc} SLED correction. The dot-dashed line indicates the best fit GRB
continuum while the solid curve shows the effect of adding a 3-parameter line
to the model.}


\clearpage

\begin{thebibliography}{}
%
\bibitem[Band et al.\ 1992]{band92}Band, D., et al. 1992, Exp. Astr., 2, 307
%
\bibitem[Band et al.\ 1993]{band93}Band, D., et al. 1993, \apj, 413, 281
%
\bibitem[Band et al.\ 1994]{band94}Band, D., et al. 1994, \apj, 434, 560
(Paper~II)
%
\bibitem[Band et al.\ 1995]{band95}Band, D., et al. 1995, \apj, 447, 289
(Paper~III)
%
\bibitem[Bevingtion 1969]{bev69}Bevington, P. R. 1969, Data Reduction and
Error Analysis for the Physical Sciences (McGraw Hill:  New York)
%
\bibitem[Briggs 1995]{brig95}Briggs, M. S. 1995, in The Annals of the New
York Academy of Sciences (17th Texas Symposium on Relativistic Astrophysics),
in press
%
\bibitem[Fishman et al.\ 1989]{fish89}Fishman, G., et al. 1989, in Proc. {\it
Gamma Ray Observatory} Science Workshop, ed. W.~N.~Johnson (Washington:
NASA),
2-39
%
\bibitem[Ford et al.\ 1994]{ford94}Ford, L., et al. 1994, in AIP Conf. 307,
Gamma-Ray Bursts, 2d Workshop, ed. G.~J.~Fishman, J.~J.~Brainerd, \& K.~Hurley
(New York: AIP), 261
%
\bibitem[Ford et al.\ 1995]{ford95}Ford, L., et al. 1995, \apj, 439, 307
%
\bibitem[Freeman \& Lamb 1994]{free94}Freeman, P. E., \& Lamb, D. Q. 1994,
Analysis of the Line Candidate in GB~930506:  Preliminary Results, University
of Chicago Internal Report
%
\bibitem[Hueter 1987]{huet87}Hueter, G. J. 1987, PhD Thesis, UC San Diego
%
\bibitem[Martin 1971]{mart71}Martin, B. R. 1971, Statistics for Physicists
(London: Academic Press)
%
\bibitem[Mazets et al.\ 1981]{Maze81}Mazets, E. P., et al. 1981, Nature, 290,
378
%
\bibitem[Meegan et al.\ 1992]{meeg92}Meegan, C. A., Fishman, G. J., Wilson, R.
B., Paciesas,~W.~S., Pendleton,~G.~N., Horack,~J.~M., Brock,~M.~N., \&
Kouveliotou,~C. 1992, Nature, 355, 143
%
\bibitem[Murakami et al.\ 1988]{mura88}Murakami, T., et al 1988, Nature, 335,
234
%
\bibitem[Palmer et al.\ 1994]{palm94}Palmer, D., et al. 1994, \apjl, 433, L77
(Paper~I)
%
\bibitem[Preece et al.\ 1994]{pree94}Preece, R., et al. 1994, in AIP Conf. 307,
Gamma Ray Bursts, 2d Workshop, ed. G.~J.~Fishman, J.~J.~Brainerd and K.~Hurley
(New York: AIP), 266
%
\bibitem[Press et al.\ 1992]{pres92}Press, W. H., Teukolsky, S.~A.,
Vetterling,~W.~T., \& Flannery,~B.~P. 1992, Numerical Recipes in FORTRAN
(2d ed.; Cambridge:  Cambridge Univ. Press)
%
\bibitem[Schaefer et al.\ 1994]{scha94}Schaefer, B. E., et al. 1994, in AIP
Conf. 307, Gamma Ray Bursts, 2d Workshop, ed. G.~J.~Fishman, J.~J.~Brainerd and
K.~Hurley (New York:  AIP), 271
%
\bibitem[Wang et al.\ 1989]{wang89}Wang, J. C. L., et al. 1989, \prl, 63, 1550

\end{thebibliography}
\end{document}